\documentclass[doublecol,linenumbers]{epl2}
\usepackage{amssymb,latexsym,amsmath,amscd}
 \title{A modified Poisson-Boltzmann theory: Effects of co-solvent polarizability}
  \shorttitle{A modified Poisson-Boltzmann equation}
\author{Yu.~A. Budkov\inst{1,2} \thanks{E-mail: \email{urabudkov@rambler.ru}} \and A. L. Kolesnikov\inst{3,4} \thanks{E-mail: \email{bancocker@mail.ru}}  \and M. G. Kiselev\inst{1}}
\shortauthor{Yu. A. Budkov \etal}

\institute{                    
  \inst{1} G.A. Krestov Institute of Solution Chemistry of the Russian Academy of Sciences, Ivanovo, Russia\\
  \inst{2} National Research University Higher School of Economics, Department of Applied Mathematics, Moscow, Russia\\
	\inst{3} Institut f\"{u}r Nichtklassische Chemie e.V., Universit\"{a}t Leipzig - Leipzig, Germany\\
	\inst{4} Ivanovo State University, Ivanovo, Russia
}

\pacs{82.45.Gj}{Electrolytes}
\pacs{33.15.Kr}{Polarizability of molecules}
\pacs{65.40.gk}{Electrochemical properties}

\abstract{In this paper within a field-theoretical approach taking into account explicitly a co-solvent with a nonzero dipole and a polarizability tensor, we derive a modified Poisson-Boltzmann equation. Applying the modified Poisson-Boltzmann equation, we formulate a generalized Gouy-Chapman theory for the case when an electrolyte solution is mixed with a polar co-solvent having a large polarizability. We show that an increase of the co-solvent concentration as well as the co-solvent polarizability lead to a significant increase of differential capacitance at sufficiently high surface potentials of the electrode, whereas the profile of the electrostatic potential becomes considerably more long-ranged. On the contrary, an increase in the permanent dipole of the co-solvent only weakly affects the differential capacitance.}

\begin{document}
\maketitle

\section{Introduction}
Up to now the Poisson-Boltzmann (PB) equation remains the simplest and the most effective theoretical tool for describing distribution of charged particles near charged surfaces in biophysics, colloid chemistry, electrochemisrty, etc. However, the PB equation contains many assumptions that make it inapplicable to real physical systems. The mean-field nature of the PB equation does not allow us to take into account the effects of the ionic correlations, whereas the consideration of the solvent as a continuous dielectric medium makes it impossible to study the effects of the solvent molecular structure. These two factors motivated the researchers to find new approaches the PB equation modifications during the last two decades \cite{Ben-Yaakov_Review,Podgornik_Review,Grochowski}.

Nowadays, great efforts have been made to modify the PB equation with respect to ionic correlations \cite{Podgornik_1989,Netz,Moreira_Netz,Netz_Orland,Forsman}, the dipole structure of the solvent \cite{Coalson_1996,Andelman_2007,Andelman_2012,Buyukdagli_2013}, polarizability and permanent dipole of ions \cite{Ben-Yaakov_2011,Frydel,Hatlo,Buyukdagli_2013} as well as their excluded volume \cite{Andelman_1997,Antypov_2005,Kornyshev,Buyukdagli,Buyukdagli_2}, and finally solvent quadrupolarizability \cite{Slavchov}. These modifications sufficiently improved the theory of electrolytes that enabled calculations of the physico-chemical characteristics of real solutions, such as dielectric permittivity and activity coefficients \cite{Andelman_2012,Slavchov}.

Nevertheless, despite the evident success in extending the pure PB-paradigm there is still no answer to the question whether or not it is possible to incorporate into the PB equation the effects of the solvent polarizability in the electric field of the ions and external charges.

Schroder and Steinhause presented \cite{Schroder_2010} a research of an influence of cation polarizability on the structure and dynamics of an ionic liquid within Molecular Dynamics (MD) simulations. It is shown that when the cation polarizability increases, the structure of ionic liquid remains almost unchanged. On the contrary, the rotational and translational diffusion sufficiently increase. Recently, using MD simulations for the set of ionic liquids was shown that cation polarizability has a significant effect on a shape of the lines of a vibration density of states \cite{Cavalcante_2014}. These observations indicate that polarizability significantly affects the characteristics that are mainly determined through the long-range correlations of the solution particles. The differential capacitance of an electric double layer is one of the physical values that is mainly determined by the long-range correlations. Thus, to correctly evaluate the differential capacitance and the electrostatic potential at long distances from the charged electrode, it should be important to incorporate the polarizability effect of the solvent and the electrolyte ions.  Moreover, simultaneous incorporation of the polarizability effects of the solvent and ions into the PB equation is especially important for electrochemical applications, where one of the main problems is the search factors that can significantly increase a magnitude of the differential capacitance.

In particular, two questions should be answered:

\begin{itemize}
\item{How much the differential capacitance will change when the polarizability of the solvent will be taken into account?}

\item{What is the difference between the effects of inclusion a permanent dipole and a polarizability of the solvent on the differential capacitance?}
\end{itemize}

Addressing these two questions, we derive within the field-theoretical approach a modified PB equation taking into account explicitly a solvent with a nonzero permanent dipole and a polarizability tensor. Using this modified PB equation, we formulate a generalized Gouy-Chapman theory \cite{Levin,Barrat_Hansen} for the case when the electrolyte solution is mixed with a polar co-solvent having a strong polarizability. We calculate the differential capacitance and the electrostatic potential profile near the charged electrode at different co-solvent concentrations. We show that adding a strong polarizable co-solvent to the electrolyte solution leads to a significant increase of the differential capacitance under the sufficiently large surface potentials of the electrode and to a more long-ranged electrostatic potential profile. We obtain that with the increase of the co-solvent polarizability, the differential capacitance dramatically increases. On the contrary, the permanent dipole of the co-solvent has only a weak effect on the value of the differential capacitance.

\section{Theory}
We consider an electrolyte solution containing $N_{+}$ point ions carrying a charge $z_{+}e$ ($z_{+}>0$), $N_{-}$ ions carrying a charge $z_{-}e$ ($z_{-}<0$), and a solvent which we shall model as a continuous dielectric medium with dielectric permittivity $\varepsilon_{s}$. In addition, we also consider $N_{c}$ molecules of a co-solvent which has a permanent dipole $p$ and a polarizability tensor $\hat{\alpha}$. Moreover, we assume that the positively charged ions (cations) have also a polarizability tensor $\hat{\alpha}_{+}$. The latter assumption covers the case when the electrolyte is an ionic liquid. The discussed system confined to a volume $V$ and has the temperature $T$. Since in this work we shall discuss the effects of polarizability and permanent dipole only, for the sake of simplicity we neglect the excluded volume of the ions as well as the co-solvent molecules. The variants of models in which the excluded volume of the ions has been taken into account have been developed in recent works \cite{Andelman_1997,Antypov_2005,Kornyshev,Buyukdagli,Buyukdagli_2}.

A canonical partition function of the above-mentioned system can be written in the following form
\begin{equation}                                                               
\label{eq:part_func}
Z=\int d\Gamma_{c}\int d\Gamma_{i}e^{-\beta H},
\end{equation}
where the following notations for the integrating measures have been introduced
\begin{equation}
\int d\Gamma_{i}(..)=\frac{\Lambda_{+}^{-3N_{+}}\Lambda_{-}^{-3N_{-}}}{N_{+}!N_{-}!}\int\limits_{V}\prod\limits_{k=1}^{N_{+}}d\bold{r}_{k}^{+}d\bold{\mathcal{P}}^{+}_{k}\int\limits_{V}  \prod\limits_{l=1}^{N_{-}}d\bold{r}_{l}^{-}(..)
\end{equation}
and
\begin{equation}
\int d\Gamma_{c}(..)=\frac{\Lambda_{c}^{-3N_{c}}}{N_{c}!}\int \prod\limits_{k=1}^{N_{c}}d\bold{p}_{k}d\bold{\mathcal{P}}_{k}\int\limits_{V} \prod\limits_{k=1}^{N_{c}}d\bold{r}_{k}^{c}(..).
\end{equation}
The Hamiltionian of the system can be written as
\begin{eqnarray}
\label{eq:hamilton}
H&=&\frac{1}{2}\sum\limits_{k=1}^{N_{c}}\mathcal{P}_{k}\hat{\bold{\alpha}}^{-1} \mathcal{P}_{k}+\frac{1}{2}\sum\limits_{l=1}^{N_{+}}\mathcal{P}^{+}_{l}\hat{\bold{\alpha}}_{+}^{-1} \mathcal{P}^{+}_{l}  \nonumber \\
&+&\frac{1}{2}\int\limits_{V}d\bold{r}\int\limits_{V}d\bold{r}^{\prime}
\hat{\rho}(\bold{r})V_{c}(\bold{r}-\bold{r}^\prime )\hat{\rho}(\bold{r}^\prime),
\end{eqnarray}
where $\mathcal{P}_{k}$ and $\mathcal{P}_{l}^{+}$ are fluctuating dipoles of $k$-th molecule of the co-solvent and $l$-th cation, respectively; $\hat{\bold{\alpha}}$ and $\hat{\bold{\alpha}}_{+}$ are the polarizability tensors of the co-solvent molecules and the cations, respectively. Moreover, a local charge density $\hat{\rho}(\bold{r})$ of the system has been introduced:
\begin{equation}
\nonumber
\hat{\rho}(\bold{r})=z_{+}e\sum\limits_{k=1}^{N_{+}}\delta\left(\bold{r}-\bold{r}^{+}_{k}\right)+z_{-}e\sum\limits_{k=1}^{N_{-}}\delta\left(\bold{r}-\bold{r}^{-}_{k}\right)
\end{equation}
\begin{equation}
\label{eq:charge_dens}
-\sum\limits_{k=1}^{N_{c}}\left(\bold{p}_{k}+\bold{\mathcal{P}}_{k}\right)\nabla \delta\left(\bold{r}-\bold{r}^{c}_{k}\right)-\sum\limits_{k=1}^{N_{+}}\bold{\mathcal{P}}^{+}_{k}\nabla \delta\left(\bold{r}-\bold{r}^{+}_{k}\right)+\rho_{ext}(\bold{r}),
\end{equation}
where $\bold{p}_{k}$ is the permanent dipole of the $k$-th co-solvent molecule; $\rho_{ext}(\bold{r})$ is the density of an external charge; $V_{c}\left(\bold{r}-\bold{r}^{\prime}\right)=\frac{1}{\varepsilon_{s}|\bold{r}-\bold{r}^{\prime}|}$ is the Coulomb potential.  It should be noted that in (\ref{eq:hamilton}) for the sake of simplicity the electrostatic self-energy has been omitted.

Using the standard Hubbard-Stratonovich transformation,  after a calculation of the Gaussian integrals over the variables $\bold{\mathcal{P}}_{j}$ and $\bold{\mathcal{P}}^{+}_{j}$ and an integration over the orientations of the dipoles $\bold{p}_{k}$\revision{,} we arrive at 
\begin{eqnarray}
Z&=&Z_{id,i}Z_{id,c}\int\frac{\mathcal{D}\varphi}{C_{0}}e^{-\frac{\varepsilon_{s}}{8\pi\beta}\int\limits_{V}d\bold{r} \left(\nabla \varphi(\bold{r})\right)^2+i\int\limits_{V}d\bold{r}\rho_{ext}(\bold{r})\varphi(\bold{r})} \nonumber \\
&\times&Q_{+}^{N_{+}}[\varphi]Q_{-}^{N_{-}}[\varphi]Q_{c}^{N_{c}}[\varphi],
\end{eqnarray}
where $C_{0}=\int\mathcal{D}\varphi e^{-\frac{\varepsilon_{s}}{8\pi\beta}\int\limits_{V}d\bold{r} \left(\nabla \varphi(\bold{r})\right)^2}$ is the normalization constant; $\beta=1/k_{B}T$;
\begin{equation}
Z_{id,i}=\frac{\Lambda_{+}^{-3N_{+}}\Lambda_{-}^{-3N_{-}}V^{N_{+}+N_{-}}}{N_{+}!N_{-}!}\left(2\pi k_{B}T\det{\hat{\alpha}_{+}}\right)^{N_{+}/2}
\end{equation}
is the ideal partition function of the ions,
\begin{equation}
Z_{id,c}=\frac{\Lambda_{c}^{-3N_{c}}V^{N_{c}}}{N_{c}!}\left(2\pi k_{B}T\det{\hat{\bold{\alpha}}}\right)^{N_{c}/2}
\end{equation}
is the ideal partition function of the co-solvent molecules; $\Lambda_{\pm}$ and $\Lambda_{c}$ are the thermal de Broglie wavelengths of the ions and the co-solvent molecules, respectively.
The single-particle partition functions of the ions and the co-solvent molecules that are immersed into an auxiliary fluctuating electric field with potential $i\varphi(\bold{r})$ take the following forms:
\begin{equation}
Q_{+}[\varphi]=\int\limits_{V}\frac{d\bold{r}}{V}e^{iz_{+}e\varphi(\bold{r})-\frac{k_{B}T}{2}\nabla{\varphi}(\bold{r})\hat{\alpha}_{+}\nabla{\varphi}(\bold{r})},
\end{equation}
\begin{equation}
Q_{-}[\varphi]=\int\limits_{V}\frac{d\bold{r}}{V}e^{iz_{-}e\varphi(\bold{r})},
\end{equation}
\begin{equation}
Q_{c}[\varphi]=\int\limits_{V}\frac{d\bold{r}}{V}e^{-\frac{k_{B}T}{2}\nabla{\varphi(\bold{r})}\hat\alpha \nabla{\varphi(\bold{r})}}\frac{\sin{p|\nabla{\varphi (\bold{r})}|}}{p|\nabla{\varphi (\bold{r})}|}.
\end{equation}

In the thermodynamic limit ($N_{\pm}\rightarrow\infty$), we have the following asymptotic relations \cite{Efimov_1996}
\begin{eqnarray}
\nonumber
Q_{-}^{N_{-}}[\varphi]&=&\left[1+\int\limits_{V}\frac{d\bold{r}}{V}\left(e^{iz_{-}e\varphi (\bold{r})}-1\right)\right]^{N_{-}} \nonumber \\
&\simeq& e^{\rho_{-}\int\limits_{V}d\bold{r}\left(e^{iz_{-}e\varphi (\bold{r})}-1\right)},
\end{eqnarray}
\begin{eqnarray}
Q_{+}^{N_{+}}[\varphi]\simeq e^{\rho_{+}\int\limits_{V}d\bold{r}\left(e^{iz_{+}e\varphi (\bold{r})-\frac{k_{B}T}{2}\nabla{\varphi(\bold{r})}\hat\alpha_{+} \nabla{\varphi(\bold{r})}}-1\right)},
\end{eqnarray}
where $\rho_{\pm}=N_{\pm}/V$ are average concentrations of the ions in the bulk solution.

Analogously, in the thermodynamic limit ($N_{c}\rightarrow\infty$) we can obtain
\begin{equation}
Q_{c}^{N_{c}}[\varphi]\simeq e^{\rho_{c}\int\limits_{V}d\bold{r}\left(e^{-\frac{k_{B}T}{2}\nabla{\varphi(\bold{r})}\hat\alpha \nabla{\varphi(\bold{r})}}\frac{\sin{p|\nabla{\varphi (\bold{r})}|}}{p|\nabla{\varphi (\bold{r})}|}-1\right)}.
\end{equation}
Thus, the excess partition function of the system takes the form
\begin{eqnarray}
\label{eq:func_int}
Z_{ex}=\frac{Z}{Z_{id,i}Z_{id,c}}=\int\frac{\mathcal{D}\varphi}{C_{0}}e^{-S[\varphi]},
\end{eqnarray}
where the following short-hand notation for the integrand has been introduced
\begin{equation}
S[\varphi]=\frac{\varepsilon_{s}}{8\pi\beta}\int\limits_{V}d\bold{r} \left(\nabla \varphi(\bold{r})\right)^2-i\int\limits_{V}d\bold{r}\rho_{ext}(\bold{r})\varphi(\bold{r})-W_{I}[\varphi],
\end{equation}
where a $"$functional of interaction$"$ $W_{I}[\varphi]$ has been also introduced
\begin{eqnarray}
\label{eq:func_inter}
W_{I}[\varphi]&=&\rho_{c}\int\limits_{V}d\bold{r}\left(e^{-\frac{k_{B}T}{2}\nabla{\varphi(\bold{r})}\hat\alpha \nabla{\varphi(\bold{r})}}\frac{\sin{ p|\nabla{\varphi (\bold{r})}|}}{p|\nabla{\varphi (\bold{r})}|}-1\right) \nonumber \\
&+& \rho_{+}\int\limits_{V}d\bold{r}\left(e^{iz_{+}e\varphi (\bold{r})-\frac{k_{B}T}{2}\nabla{\varphi(\bold{r})}\hat\alpha_{+} \nabla{\varphi(\bold{r})}}-1\right)\nonumber \\
&+&\rho_{-}\int\limits_{V}d\bold{r}\left(e^{iz_{-}e\varphi (\bold{r})}-1\right).
\end{eqnarray}
Then, we calculate the functional integral (\ref{eq:func_int}) within a standard saddle-point approximation \cite{Podgornik_Review}. An equation for the saddle-point has the form
\begin{equation}
\label{eq:saddle-point}
\frac{\delta S[\varphi]}{\delta \varphi(\bold{r})}=0.
\end{equation}
Calculating the variational derivative in (\ref{eq:saddle-point}) and introducing a notation for an electrostatic potential $\psi(\bold{r})=-ik_{B}T\varphi(\bold{r})$, one can obtain the following equation
\begin{eqnarray}
\nabla\left(\hat{\varepsilon}(\bold{r})\nabla{\psi}(\bold{r})\right)=-4\pi\rho_{ext}(\bold{r})-4\pi z_{-}e\rho_{-}e^{-\frac{z_{-}e\psi(\bold{r})}{k_{B}T}}\nonumber
\end{eqnarray}
\begin{eqnarray}
\label{eq:poisson-boltz_eq}
-4\pi z_{+}e\rho_{+}e^{-\frac{z_{+}e\psi(\bold{r})}{k_{B}T}+\frac{\nabla{\psi(\bold{r})}\hat\alpha_{+} \nabla{\psi(\bold{r})}}{2k_{B}T}},
\end{eqnarray}
where a tensor of the local dielectric permittivity has been introduced
\begin{eqnarray}
\label{eq:dielectr_perm}
\hat{\varepsilon}(\bold{r})&=&\varepsilon_{s}\bold{I}+4\pi \rho_{c} e^{\frac{\nabla{\psi(\bold{r})}\hat\alpha \nabla{\psi(\bold{r})}}{2k_{B}T}}\frac{\sinh{\beta p|\nabla{\psi}(\bold{r})|}}{\beta p|\nabla{\psi}(\bold{r})|} \nonumber \\
&\times& \left(\hat{\bold{\alpha}}+\bold{I}\frac{p^2}{k_{B}T}\frac{L(\beta p|\nabla{\psi}(\bold{r})|)}{\beta p|\nabla{\psi}(\bold{r})|}\right) \nonumber \\
&+&4\pi \rho_{+}\hat{\alpha}_{+}e^{-\frac{z_{+}e\psi(\bold{r})}{k_{B}T}+\frac{\nabla{\psi(\bold{r})}\hat\alpha_{+} \nabla{\psi(\bold{r})}}{2k_{B}T}},
\end{eqnarray}
where
$L(x)=\coth{x}-1/x$ is a Langevin function. The tensor of dielectric permittivity in the bulk solution (where $\psi=0$) is determined by the following expression
\begin{equation}
\hat{\varepsilon}=\varepsilon_{s}\bold{I}+4\pi\rho_{c}\left(\hat{\bold{\alpha}} +\bold{I}\frac{p^2}{3k_{B}T}\right)+ 4\pi\rho_{+}\hat{\alpha}_{+},
\end{equation}
where $\bold{I}$ is the identity tensor; $\rho_{c}=N_{c}/V$ is a bulk co-solvent concentration.

The equations (\ref{eq:poisson-boltz_eq}-\ref{eq:dielectr_perm}) determine the electrostatic potential $\psi(\bold{r})$ and the tensor of local dielectric permittivity within a self-consistent field approximation. The latter is a generalization of the classical PB theory for the case of an explicit account of the polar polarizable co-solvent. If one assumes that $\hat{\bold{\alpha}}=\hat{\bold{\alpha}}_{+}=0$ and $\varepsilon_{s}=1$, then one arrives at the expression which is obtained in the work \cite{Andelman_2012}
\begin{equation}
\label{eq:poisson-boltz_eq2}
\varepsilon(\bold{r})=1+\frac{4\pi \rho_{c}p^2}{k_{B}T}\mathcal{G}(\beta p|\nabla{\psi}(\bold{r})|),
\end{equation}
where $\mathcal{G}(x)=\frac{\sinh{x}}{x^2}L(x)$.

In the case of nonpolar co-solvent ($p=0$), we obtain the following expression for the tensor of local dielectric permittivity:
\begin{eqnarray}
\hat{\varepsilon}(\bold{r})&=&\varepsilon_{s}\bold{I}+4\pi \rho_{c}\hat{\bold{\alpha}} e^{\frac{\nabla{\psi(\bold{r})}\hat\alpha \nabla{\psi(\bold{r})}}{2k_{B}T}} \nonumber \\
&+&4\pi \rho_{+}\hat{\alpha}_{+} e^{-\frac{z_{+}e\psi(\bold{r})}{k_{B}T}+\frac{\nabla{\psi(\bold{r})}\hat\alpha_{+} \nabla{\psi(\bold{r})}}{2k_{B}T}}.
\end{eqnarray}

Finally, an excess electrostatic free energy of the solution can be written in the following form
\begin{eqnarray}
&F_{el}[\psi]&=-\frac{\varepsilon_{s}}{8\pi}\int\limits_{V}d\bold{r}(\nabla{\psi}(\bold{r}))^2 \nonumber \\
&-&\rho_{-}k_{B}T\int\limits_{V}d\bold{r}\left(e^{-\frac{z_{-}e\psi(\bold{r})}{k_{B}T}}-1\right) \nonumber \\
&-&\rho_{+}k_{B}T\int\limits_{V}d\bold{r}\left(e^{-\frac{z_{+}e\psi(\bold{r})}{k_{B}T}+\frac{\nabla{\psi(\bold{r})}\hat\alpha_{+} \nabla{\psi(\bold{r})}}{2k_{B}T}}-1\right) \nonumber \\
&-&\rho_{c}k_{B}T\int\limits_{V}d\bold{r}\left(e^{\frac{\nabla{\psi(\bold{r})}\hat\alpha \nabla{\psi(\bold{r})}}{2k_{B}T}}\frac{\sinh{\beta p|\nabla{\psi (\bold{r})}|}}{\beta p|\nabla{\psi (\bold{r})}|}-1\right) \nonumber \\
&+&\int\limits_{V}d\bold{r}\rho_{ext}(\bold{r})\psi(\bold{r}).
\end{eqnarray}

\section{The differential capacitance of the electric double layer: Effect of co-solvent polarizability and permanent dipole}
As an illustration of the application of the modified PB equation (\ref{eq:poisson-boltz_eq}-\ref{eq:dielectr_perm}), we formulate the generalized Gouy-Chapman theory \cite{Levin,Barrat_Hansen}. We consider a system containing a charged electrode, which we shall model as a charged flat surface with a fixed surface charge density $\sigma$, the point ions of 1:1 electrolyte ($z_{+}=-z_{-}=1$), and the point molecules of the polar polarizable co-solvent with a polarizability $\hat{\bold{\alpha}}=\alpha \bold{I}$ and a permanent dipole moment $p$. In this case the average concentrations of ions in the bulk are equal, i.e.  $\rho_{+}=\rho_{-}=\rho$. For the sake of simplicity, we assume that $\hat{\alpha}_{+}=0$.

Choosing $z$ axis perpendicular to the electrode and placing the origin  on it, one can write the modified PB equation in the following form
\begin{equation}
\label{eq:poisson-boltz_eq4}
\frac{d}{dz}\left(\varepsilon(z)\psi^{\prime}(z)\right)=8\pi\rho e \sinh{\beta e \psi (z)},
\end{equation}
where the effective dielectric permittivity $\varepsilon(z)$ takes the form
\begin{eqnarray}
\nonumber
\label{eq:permittivity}
\varepsilon(z)=\varepsilon_{s}+4\pi\rho_{c}e^{\frac{\alpha \psi^{\prime}(z)^2}{2 k_{B}T}}\frac{\sinh{\beta p\psi^{\prime}(z)}}{\beta p\psi^{\prime}(z)}
\end{eqnarray}

\begin{eqnarray}
\times\left(\alpha+\frac{p^2}{k_{B}T}\frac{L(\beta p \psi^{\prime}(z))}{\beta p \psi^{\prime}(z)}\right).
\end{eqnarray}
In order to obtain the electrostatic potential profile $\psi(z)$, we have to formulate the standard boundary condition \cite{Barrat_Hansen,Landau_VIII}

\begin{eqnarray}
\label{eq:boundary_cond}
-\varepsilon(0)\psi^{\prime}(0)=4\pi\sigma.
\end{eqnarray}

Using the standard method of integrating the equations of this form, taking into account the second boundary condition $\psi^{\prime}(\infty)=0$, we can obtain the first integral of equation (\ref{eq:poisson-boltz_eq4})
\begin{equation}
\label{eq:first_int}
\frac{\varepsilon_{s}\mathcal{E}^2(z)}{8\pi}+\int\limits_{0}^{\mathcal{E}(z)}D(\mathcal{E}^{\prime})d\mathcal{E}^{\prime}=2\rho k_{B}T \left(\cosh(\beta e \psi (z))-1\right),
\end{equation}
where $\mathcal{E}(z)=-\psi^{\prime}(z)$ is the electric field and the auxiliary function $D(\mathcal{E})$ takes the form
\begin{equation}
\nonumber
see~eq.~(\ref{eq:D})~below
\end{equation}

\begin{widetext}
\begin{equation}
\label{eq:D}
D(\mathcal{E})=\rho_{c}e^{\frac{\alpha \mathcal{E}^2}{2k_{B}T}}\frac{\sinh{\beta p\mathcal{E}}}{\beta p}\left(\left(\alpha+\frac{p^2}{k_{B}T}\frac{L(\beta p \mathcal{E})}{\beta p \mathcal{E}}\right)\left(\frac{\alpha\mathcal{E}^2}{k_{B}T}+\beta p\mathcal{E}\coth{\beta p\mathcal{E}}\right)+\frac{p^2}{k_{B}T}\left(L^{\prime}(\beta p \mathcal{E})-\frac{L(\beta p \mathcal{E})}{\beta p \mathcal{E}}\right)\right).
\end{equation}
\end{widetext}

In the case of a nonpolar solvent ($p=0$), the first integral (\ref{eq:first_int}) can be expressed through the elementary functions as
\begin{eqnarray}
\label{eq:first_int2}
\frac{\varepsilon_{s}\mathcal{E}^2(z)}{8\pi}&+&\rho_{c}k_{B}T\left(1-e^{\frac{\alpha \mathcal{E}^2(z)}{2k_{B}T}}\right)+\rho_{c}\alpha \mathcal{E}^2(z)e^{\frac{\alpha \mathcal{E}^2(z)}{2k_{B}T}} \nonumber \\
&=&2\rho k_{B}T\left(\cosh{\beta e\psi(z)}-1\right).
\end{eqnarray}

To integrate the equation (\ref{eq:first_int}), we should first solve it as a transcendental equation with respect to $\mathcal{E}=-\psi^{\prime}(z)$ at different values $\psi$.  Thus, we obtain a function $\mathcal{E}=\mathcal{E}(\psi)$. The next step consists of solving the equation $\psi^{\prime}=-\mathcal{E}(\psi)$.

Before calculating the electrostatic potential profile $\psi(z)$, we estimate the differential capacitance $C$. As it was already pointed out in Introduction, this quantity should be sensitive to an account of the co-solvent polarizability. The differential capacitance can be determined by the following relation
\begin{equation}
C=\frac{d\sigma}{d\psi_{0}},
\end{equation}
where $\psi_{0}=\psi(0)$ is the surface potential of the electrode, which in contrast to the surface charge density $\sigma$ is usually an experimentally controllable parameter.

To perform the numerical calculations, we introduce dimensionless parameters in the following way
\begin{equation}
\label{eq:dimensionless_units}
\tilde{p}=p/el_{B}, ~~ \tilde{\alpha} =\alpha/l_{B}^3\varepsilon_{s}, ~~ \theta=l_{B}^3\rho,~~ \gamma =l_{B}^3\rho_{c},
\end{equation}
where $l_{B}=e^2/\varepsilon_{s}k_{B}T$ is the Bjerrum length. Moreover, we introduce the dimensionless electrostatic potential $u=\beta e\psi$, the electric field $\tilde{\mathcal{E}}=\mathcal{E}\beta e l_{B}$, and the distance from the electrode $\tilde{z}=z/l_{B}$. 
We also introduce the dimensionless surface charge density $\tilde{\sigma}=\sigma \beta el_{B}/\varepsilon_{s}$, which is related to the dimensionless electric field at the electrode $\tilde{\mathcal{E}}_{0}=\tilde{\mathcal{E}}(0)$ via the boundary condition (\ref{eq:boundary_cond}). The latter can be rewritten in the dimensionless form as
\begin{equation}
\label{eq:boundary_cond_2}
\left(1+4\pi\gamma e^{\frac{\tilde{\alpha}\tilde{\mathcal{E}_{0}^2}}{2}}\xi(\tilde{\mathcal{E}}_{0}) \right)\tilde{\mathcal{E}}_{0}=4\pi \tilde{\sigma},
\end{equation}
where $\xi(\tilde{\mathcal{E}}_{0})=\frac{\sinh{\tilde{p}\tilde{\mathcal{E}}_{0}}}{\tilde{p}\tilde{\mathcal{E}}_{0}}\left(\tilde{\alpha}+\tilde{p}\frac{L(\tilde{p}\tilde{\mathcal{E}}_{0})}{\tilde{\mathcal{E}}_{0}}\right)$.
\begin{figure}
\centerline{\includegraphics[scale = 0.25]{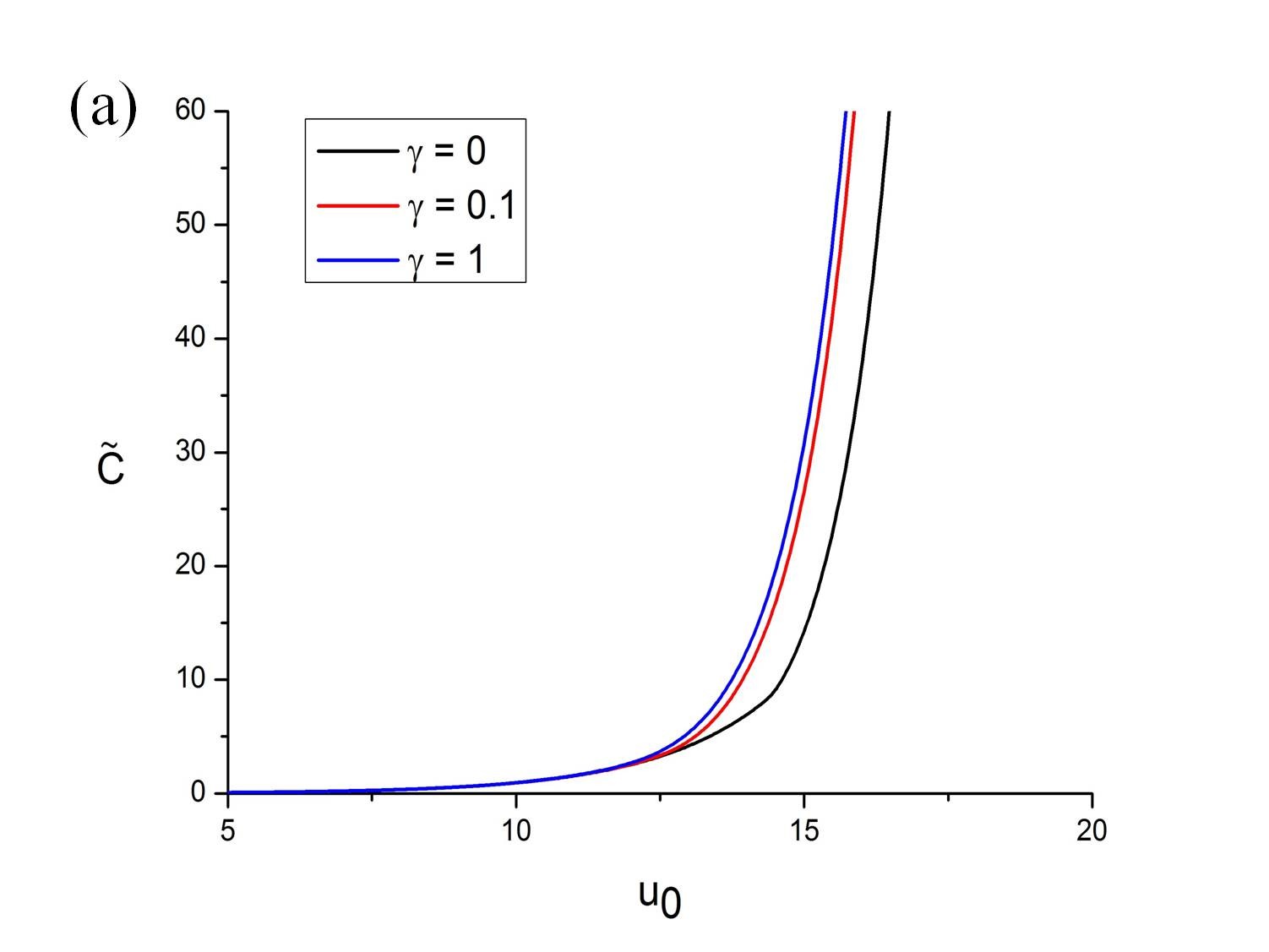}}
\centerline{\includegraphics[scale = 0.25]{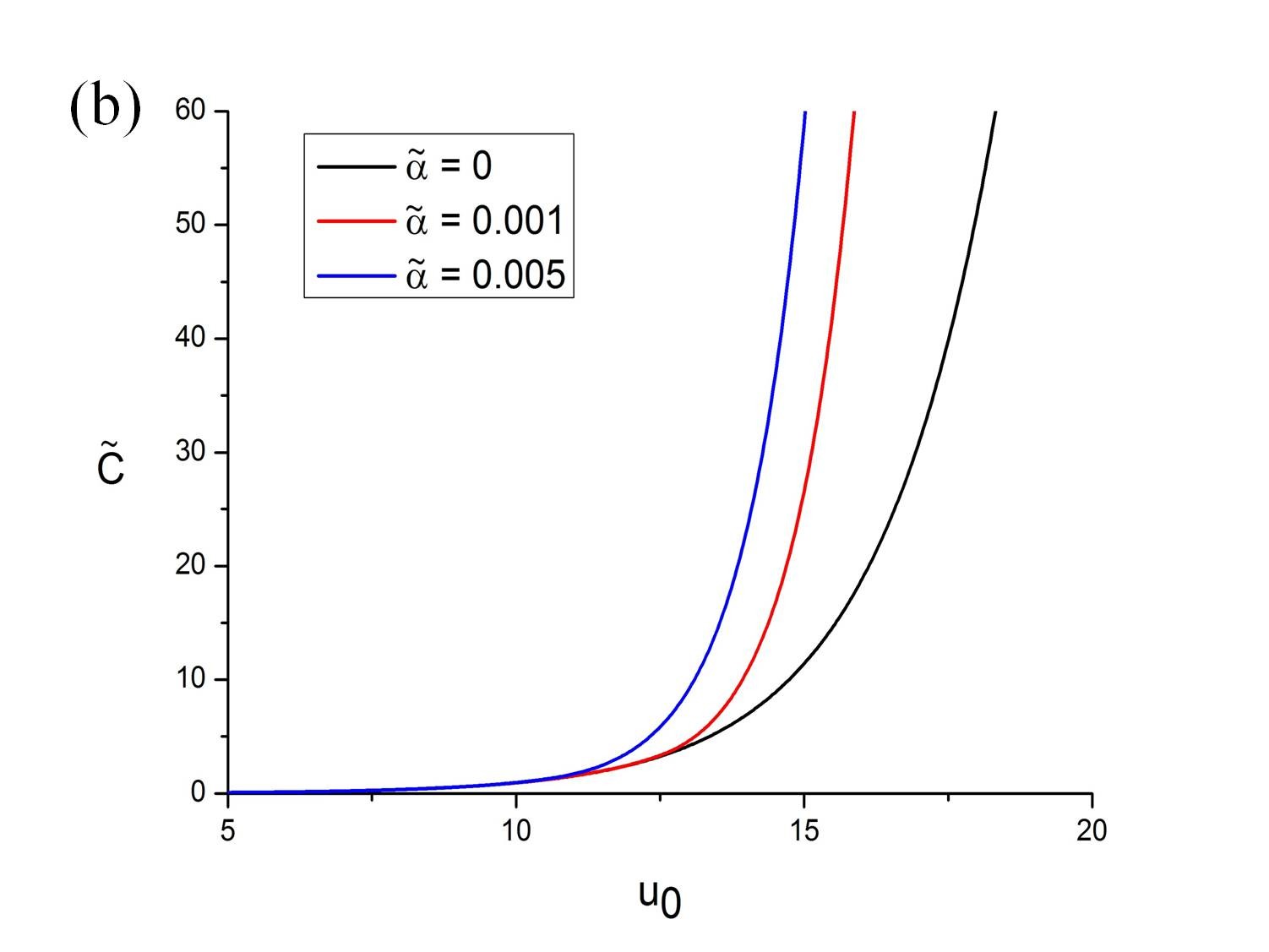}}
\caption{(a) The differential capacitance $\tilde{C}$ as a function of the surface potential $u_{0}$ at different $\gamma$ and at fixed parameters $\tilde{\alpha}=10^{-3}$, $\theta=10^{-3}$ and $\tilde{p}=0$. (b) The differential  capacitance as a function of the surface potential $u_{0}$ at the different co-solvent polarizability $\tilde{\alpha}$ and at fixed $\gamma=0.1$, $\theta=10^{-3}$ and $\tilde{p}=0$.}
\label{fig.1}
\end{figure}

Further, using the definition of the reduced differential capacitance
\begin{equation}
\label{eq:dimensionless_capac}
\tilde{C}=\frac{d\tilde{\sigma}}{du_{0}},
\end{equation}
equation (\ref{eq:first_int}) at $\tilde{z}=0$ is written in the dimensionless form

\begin{equation}
\label{eq:E_vs_u}
\tilde{\mathcal{E}}_{0}^2+8\pi\int\limits_{0}^{\tilde{\mathcal{E}}_{0}}dv\tilde{D}\left(v\right)=16\pi \theta\left(\cosh{u_{0}}-1\right),
\end{equation}
after some algebra, one can obtain
\begin{equation}
\label{eq:capacitance_final}
\tilde{C}=\frac{2\theta \sinh{u_{0}}}{\tilde{\mathcal{E}}_{0}(u_{0})},
\end{equation}
where the function $\tilde{\mathcal{E}}_{0}=\tilde{\mathcal{E}}_{0}(u_{0})$ is determined implicitly through (\ref{eq:E_vs_u}) and $u_{0}=u(0)$. The dimensionless auxiliary function $\tilde{D}\left(v\right)$ has the following form
\begin{equation}
\nonumber
see~ eq.~(\ref{eq:D2})~below
\end{equation}

\begin{widetext}
\begin{equation}
\label{eq:D2}
\tilde{D}(v)=\gamma e^{\frac{\tilde{\alpha}v^2}{2}}\frac{\sinh{\tilde{p}v}}{\tilde{p}}\left(\left(\tilde{\alpha}+\tilde{p}\frac{L(\tilde{p}v)}{v}\right)\left(\tilde{\alpha}v^2+\tilde{p}v\coth{\tilde{p}v}\right)+\tilde{p}^2\left(L^{\prime}(\tilde{p} v)-\frac{L(\tilde{p} v)}{\tilde{p} v}\right)\right).
\end{equation}
\end{widetext}

At $\gamma\rightarrow 0$ we obtain $\mathcal{\tilde{E}}_{0}\simeq 4\sqrt{2\pi\theta}\sinh\frac{u_{0}}{2}$. In this case, we arrive at the well-known result obtained by Grahame \cite{Grahame} in the framework of the standard Gouy-Chapman model
\begin{equation}
\tilde{C}=\sqrt{\frac{\theta}{2\pi}}\cosh{\frac{u_{0}}{2}}.
\end{equation}

\begin{figure}
\centerline{\includegraphics[scale = 0.22]{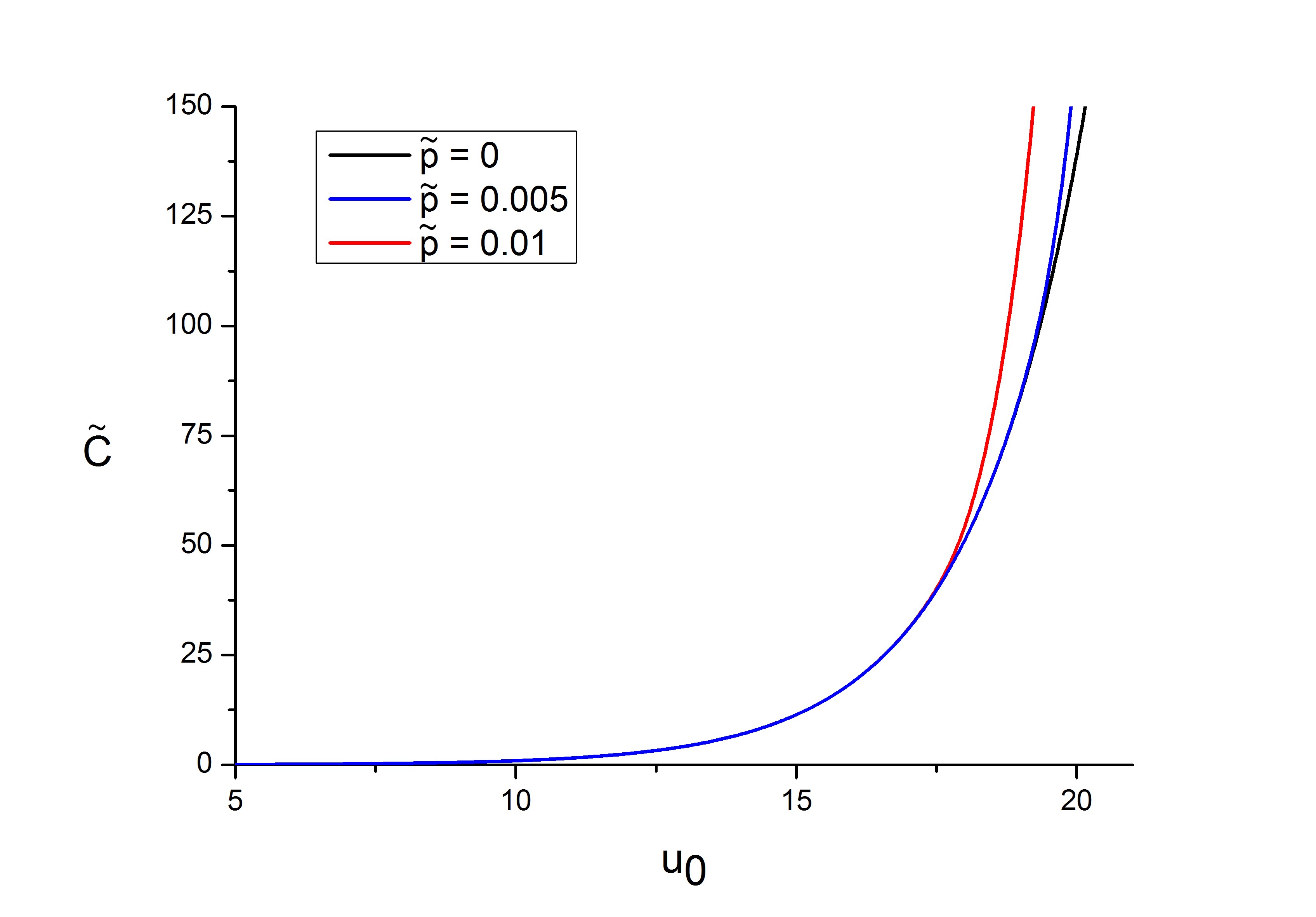}}
\caption{The dependences of the differential capacitance $\tilde{C}$ on the surface potential $u_{0}$ at different permanent dipole $\tilde{p}$ of the co-solvent. The data are shown for $\gamma=0.1$, $\theta=10^{-3}$, $\tilde{\alpha}=0$ .}
\label{fig.2}
\end{figure}

Fig 1a shows the differential capacitance as a function of the surface potential at different values of parameter $\gamma$ and fixed values $\tilde{\alpha}=10^{-3}$, $\theta=10^{-3}$, and $\tilde{p}=0$ which approximately correspond to the real systems (such as benzene or toluene, dissolved in a dilute aqueous electrolyte solution). As it can be seen from Fig. 1, the differential capacitance does not almost depend on the co-solvent concentration at the small values of $u_{0}$. However, adding the polarizable co-solvent at sufficiently large surface potential leads to a significant increase in the differential capacitance. The same trend takes place at increasing the dimensionless co-solvent polarizability (see, Fig. 1b). Fig.2 demonstrates the dependences of the differential capacitance on the surface potential at different values of the dimensionless permanent dipole $\tilde{p}$ of the co-solvent molecules. In contrast to the previous case, when the polarizability increased, an increase of the co-solvent permanent dipole leads only to a slight increase of the differential capacitance. The dependences of the electrostatic potential profiles $u(\tilde{z})$ at different $\gamma$ are marked in Fig. 4. As it can be seen, the presence of a polarizable co-solvent in the electric double layer makes the electrostatic potential considerably more long-ranged.

\begin{figure}
\centerline{\includegraphics[scale = 0.22]{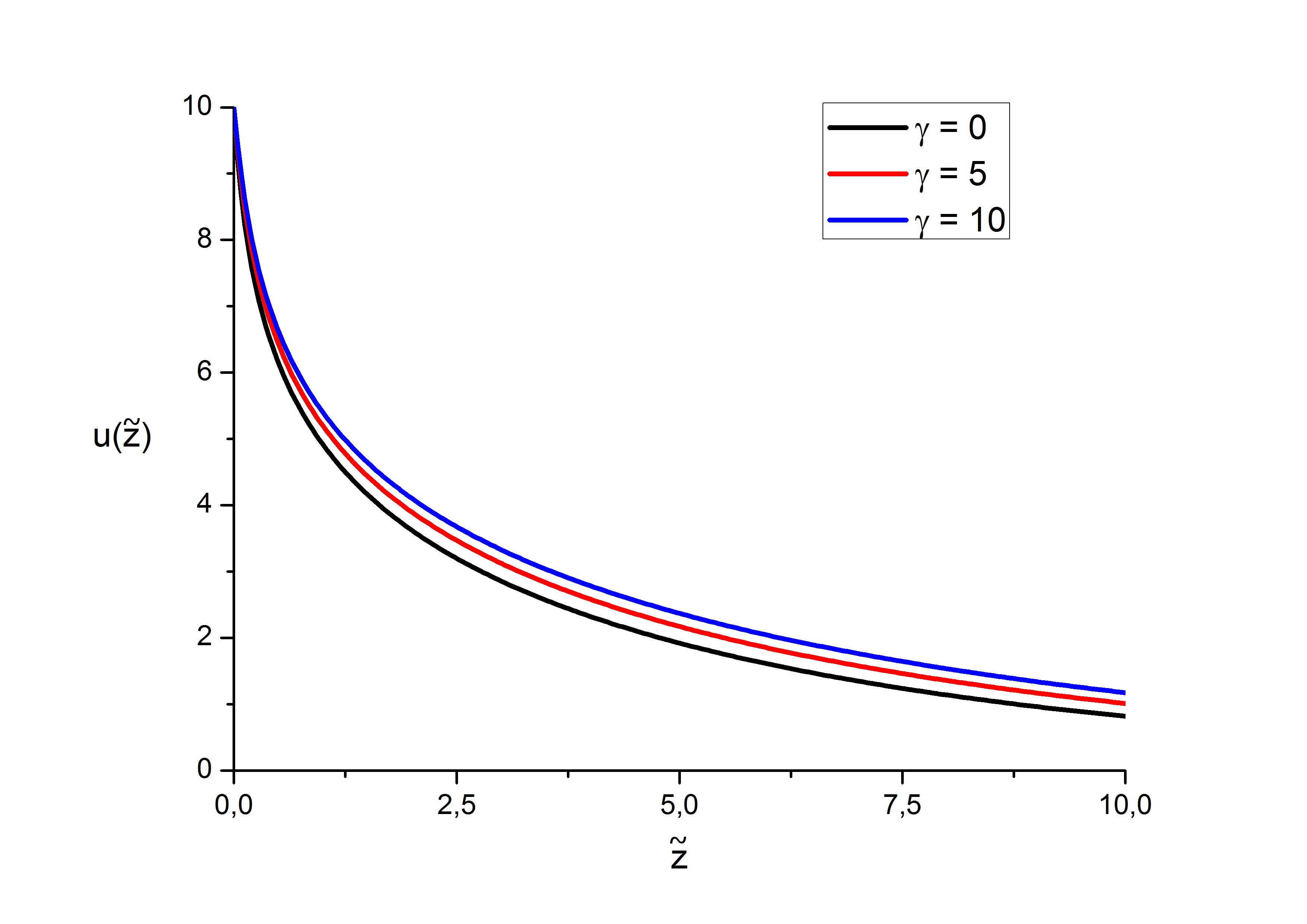}}
\caption{The electrostatic potential profiles $u(\tilde{z})$ at different values of the parameter $\gamma$. With the increase of co-solvent concentration the electrostatic potential becomes substantially more long-ranged. The data are shown for $\tilde{\alpha}=0.005$, $\theta=10^{-3}$, $u_{0}=10$, $\tilde{p}=0$.}
\label{fig.4}
\end{figure}

\section{Summary}
In this letter, in the framework of a field-theoretical formalism we have derived a modified Poisson-Boltzmann equation for the case of an explicit account of a polarizable polar co-solvent. Moreover, an expression for the tensor of the local dielectric permittivity has been obtained. 
As an important application of the developed mean-field theory, we have formulated a generalized Gouy-Chapman model of the flat electric double layer with the explicit account of the added polar polarizable co-solvent. We have shown that adding the strong polarizable co-solvent leads to a significant increase of the differential capacitance in a region of the sufficiently high surface potentials of the electrode. We have also shown that increasing the co-solvent polarizability leads to a significant increase in the differential capacitance, while growth of the permanent dipole gives only a slight enhancement the latter. As it follows from (\ref{eq:E_vs_u}--\ref{eq:capacitance_final}) the differential capacitance is proportional to the value of the local dielectric permittivity at the electrode. Thus, it should be sensitive to a change of the co-solvent concentration as well as the co-solvent polarizability (see, (\ref{eq:permittivity})). It explains such dramatic behavior of the differential capacitance when these variables increase. In addition, we have shown that the presence of strongly polarizable co-solvent molecules in the diffuse layer (at the large values of surface potential) leads to a more long-ranged electrostatic potential profile. This effect results from the fact that with an increase of the local co-solvent concentration in the diffuse layer the local dielectric permittivity increases. The latter leads to a decrease of the electrode charge screening.

In conclusion, we would like to discuss the weak point of the present theory and its possible generalization. First, our model should be valid at the  low co-solvent concentration only. Second, within the present study we have neglected the excluded volume interactions for the ions as well as for the co-solvent molecules. Nevertheless, as Kornyshev clearly showed in the work \cite{Kornyshev}, the excluded volume interactions should be also important in the region of large surface potential for a correct evaluation of the differential capacitance. Moreover, it is evident in advance the account of the excluded volume interactions for the co-solvent should decrease the discussed effect of the increase of the differential capacitance. However, we believe that the latter is still significant even taking into account the excluded volume of the co-solvent molecules as well as the electrolyte ions. A discussion of the importance of the excluded volume contributions together with the polarizability and permanent dipole of species on the properties of the electric double layer is a subject of the future publications.

\section{Acknoledgements}
Yu.A.B. thanks Prof. N.V. Brilliantov for valuable discussions that helped us to improve this work. We thank reviewers for valuable comments and remarks. This work was supported by grant from the President of RF (Grant No. MK-2823.2015.3).


\begin{thebibliography}{99}
\bibitem{Ben-Yaakov_Review}
{Dan Ben-Yaakov, David Andelman, Daniel Harries and Rudi Podgornik} J.Phys.: Condens. Matter. $\bold{21}$ (2009) 424106.

\bibitem{Podgornik_Review}
{Naji A., Kanduc M., Forsman J., Podgornik R.} J. Chem. Phys. $\bold{139}$ (2013) 150901.

\bibitem{Grochowski}
{Grochowski P., Trylska J.} Biopolymers $\bold{89}$ (2008) 93.

\bibitem{Netz}
{Netz R.R.} Eur. Phys. J. E $\bold{5}$ (2001) 557.

\bibitem{Moreira_Netz}
{A. G. Moreira and R. R. Netz} Europhys. Lett. $\bold{52}$ (2000) 705.

\bibitem{Podgornik_1989}
{Rudi Podgornik} J. Chem. Phys. $\bold{91}$ (1989) 5840.

\bibitem{Netz_Orland}
{R.R. Netz, H. Orland} The European Physical Journal E $\bold{2-3}$ (2000) 203.

\bibitem{Forsman}
{Forsman J.} J. Phys. Chem. B $\bold{108}$ (2004) 9236.

\bibitem{Coalson_1996}
{Rob D. Coalson, A. Duncan and N. B. Tal} J. Phys. Chem. $\bold{100}$ (1996) 2612.

\bibitem{Andelman_2007}
{Abrashkin A., Andelman D., Orland H.} PRL $\bold{99}$ (2007) 077801.

\bibitem{Andelman_2012}
{Levy A., Andelman D., Orland H.} PRL $\bold{108}$ (2012) 227801.

\bibitem{Buyukdagli_2014}
{Sahin Buyukdagli and Ralf Blossey} J. Chem. Phys. $\bold{140}$ (2014) 234903.

\bibitem{Ben-Yaakov_2011}
{Dan Ben-Yaakov, David Andelman, and Rudi Podgornik} J. Chem. Phys. $\bold{134}$ (2011) 074705.

\bibitem{Frydel}
{Frydel D.} J. Chem. Phys. $\bold{134}$ (2011) 234704.

\bibitem{Hatlo}
{M. M. Hatlo, R. van Roij and L. Lue} EPL $\bold{97}$ (2012) 28010.

\bibitem{Buyukdagli_2013}
{S. Buyukdagli and T. Ala-Nissila} Phys. Rev. E  $\bold{87}$ (2013) 063201.

\bibitem{Andelman_1997}
{Borukhov I., Andelman D., Orland H.} PRL $\bold{79}$ (1997) 435.

\bibitem{Antypov_2005}
{Dmytro Antypov, Marcia C. Barbosa, Christian Holm} PRE $\bold{71}$ (2005) 061106.

\bibitem{Kornyshev}
{Kornyshev A.} J. Phys. Chem. B $\bold{111}$ (2007) 5545.

\bibitem{Buyukdagli}
{S. Buyukdagli and T. Ala-Nissila} EPL $\bold{98}$ (2012) 60003.

\bibitem{Buyukdagli_2}
{S. Buyukdagli, C.V. Achim and T. Ala-Nissila} J. Stat. Mech. $\bold{P05033}$ (2011) 1.

\bibitem{Slavchov}
{Slavchov R.I.} J. Chem. Phys. $\bold{140}$ (2014) 164510.

\bibitem{Levin}
{Levin Y.} Rep. Prog. Phys. $\bold{65}$ (2002) 1577.

\bibitem{Barrat_Hansen}
{Barrat J.-L., Hansen J.-P.} {\sl Basic concepts for simple and complex liquids} (University Press, Cambridge) 2003.

\bibitem{Schroder_2010}
{Christian Schroder and Othmar Steinhause} J. Chem. Phys. $\bold{142}$ (2015) 064503.

\bibitem{Cavalcante_2014}
{Ary de Oliveira Cavalcant, Mauro C. C. Ribeir and Munir S. Skaf} J. Chem. Phys. $\bold{140}$ (2014) 144108.

\bibitem{Efimov_1996}
{Efimov G.V., Nogovitsin E.A.} Physica A  $\bold{234}$ (1996) 506.

\bibitem{Landau_VIII}
{\sl Landau L.D., Lifshitz E.M.} {\sl Electrodynamics of Continuous Media V. 8 of A Course of Theoretical Physics}, (Pergamon Press, Oxford, UK) 1960.

\bibitem{Grahame}
{Grahame S.L.} J. Chem. Phys. $\bold{18}$ (1950) 903.




\end{thebibliography}
\end{document}